\pdfoutput=1
\documentclass[twocolumn,preprintnumbers,amsmath,amssymb]{revtex4}
\usepackage{graphicx}
\usepackage{dcolumn}
\usepackage{bm}
\begin{document}
\title{Direct observation of nanometer-scale pinning sites in\\ (Nd$_{0.33}$,Eu$_{0.20}$,Gd$_{0.47}$)Ba$_2$Cu$_3$O$_{7-\delta}$ single crystals}
\author{P. Das}
\altaffiliation[Present address: Max Planck Institute for Chemical Physics of Solids, Noethnitzer Str. 40, D-01187 Dresden, Germany ]{}
 \email{Pintu.Das@cpfs.mpg.de}
\author{M. R. Koblischka}
\affiliation{ Institute of Experimental Physics, Saarland
University, D-66041 Saarbruecken, Germany}
\author{S. Turner}
\author{G. Van Tendeloo}
\affiliation{EMAT Research Group, University of Antwerp,
Groeneborgerlaan 171, B-2020 Antwerpen, Belgium}
\author{Th. Wolf}
\affiliation{ Forschungszentrum Karlsruhe GmbH, Institute of Solid
State Physics, D-76021 Karlsruhe, Germany}
\author{M. Jirsa}
\affiliation{Institute of Physics ASCR, Na Slovance 2, CZ-182 21
Prague 8, Czech Republic}
\author{U. Hartmann}
  \affiliation{Institute of Experimental Physics, Saarland University, D-66041 Saarbruecken, Germany}

\begin{abstract}
We report on the observation of self-organized stripe-like
structures on the as-grown surface and in the bulk of
(Nd,Eu,Gd)Ba$_2$Cu$_3$O$_y$ single crystals. The periodicity of the
stripes on the surface lies between 500-800 nm. These are possibly
the growth steps of the crystal. Transmission electron microscopy
investigations revealed stripes of periodicity in the range of 20-40
nm in the bulk. From electron back scattered diffraction
investigations, no crystallographic misorientation due to the
nanostripes has been found. Scanning tunneling spectroscopic
experiments revealed nonsuperconducting regions, running along twin
directions, which presumably constitute strong pinning sites.
\end{abstract}

\maketitle

The practical usefulness of a superconductor depends on various
properties, especially on the critical current density ($J_c$) and
irreversibility field ($H_{\rm{irr}}$). The high-temperature
superconductors (HTSC) are important for future applications because
the superconducting properties of these materials can be utilized at
the boiling temperature of liquid nitrogen~\cite{Murakami1996}. At
higher temperatures, due to thermally activated depinning, flux
motion appears, which is associated with an increase of electrical
resistivity and energy losses~\cite{Blatter1994}. Therefore, the
efficient pinning of vortices at elevated temperatures is the key
challenge for HTSC cuprates. For strong pinning, defects with a
diameter comparable to 2$\xi(T)$ are necessary, where $\xi$ is the
coherence length~\cite{Takezawa1997, Sosnowski2006}. Although a
normal conducting defect provides the best pinning, superconducting
inclusions with modified thermodynamic properties have also been
found to act as effective pinning sites, especially at elevated
temperatures and/or fields~\cite{Blatter1994}. By introducing
secondary-phase, nonstoichiometric inclusions in the melt-processed
samples, Murakami \textit{et al.} achieved a significant increase of
$J_c$ in YBa$_2$Cu$_3$O$_y$ (Y-123)~\cite{Murakami1989}. Even higher
$J_c$ values have been obtained in light rare earth (LRE) compounds,
especially in the ternary (Nd,Eu,Gd)-123, (Sm,Eu,Gd)-123
ones~\cite{Muralidhar2003, Hu2005, Muralidhar2005}.
\begin{figure}
\begin{center}
\includegraphics[width=0.5\textwidth]{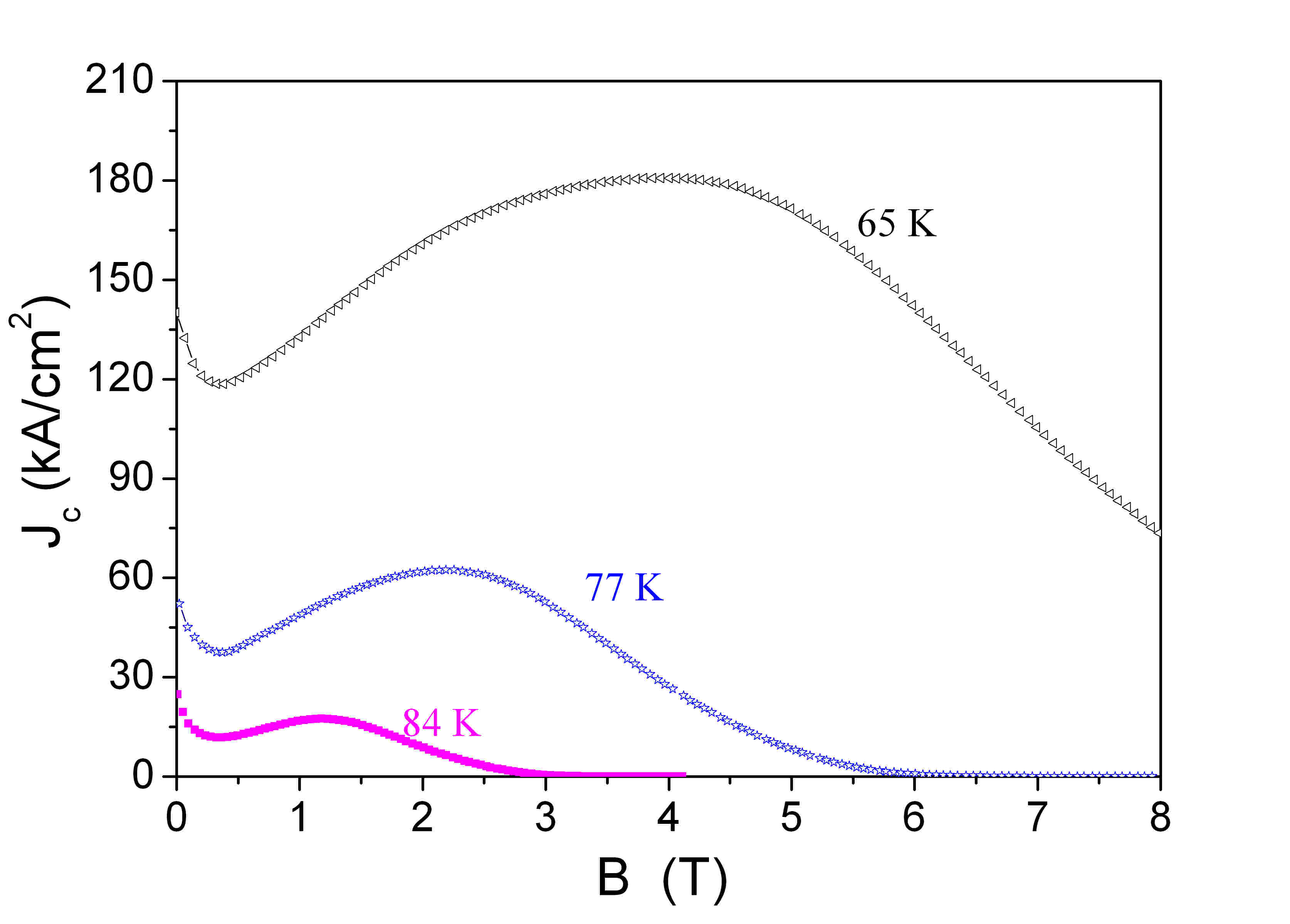}
\caption{$J_c$ as calculated from magnetization measurements at $T$=
64 K, 77 K and 84 K plotted against applied field. \label{FigJc}}
\end{center}
\end{figure}
Recently, microstructure analysis of the melt-processed ternary
LRE-123 samples showed the presence of lamellar structures of
periodicity ranging from a few nm to a few tens of nm
\cite{Muralidhar2003, Hu2005, Koblischka2007}. The samples showed
local composition fluctuations at the nanometer scale. Single
crystals, which generally offer a high morphological quality, are
important for such investigations.

In this work, we report on microstructure studies on single crystals
of Gd-enriched
(Nd$_{0.33}$,Eu$_{0.20}$,Gd$_{0.47}$)Ba$_2$Cu$_3$O$_{7-\delta}$
(NEG-123) using atomic force microscopy (AFM), scanning tunneling
microscopy (STM), transmission electron microscopy (TEM) and
electron back scattered diffraction (EBSD) techniques. The goal of
this investigation is to identify the potential pinning sites
originated from crystal growth. A Gd-rich NEG-123 sample was chosen
for this study as melt-processed samples of this type have shown
pronounced stripe-like structures~\cite{Koblischka2006}.
\begin{figure}
\includegraphics[width=0.4\textwidth]{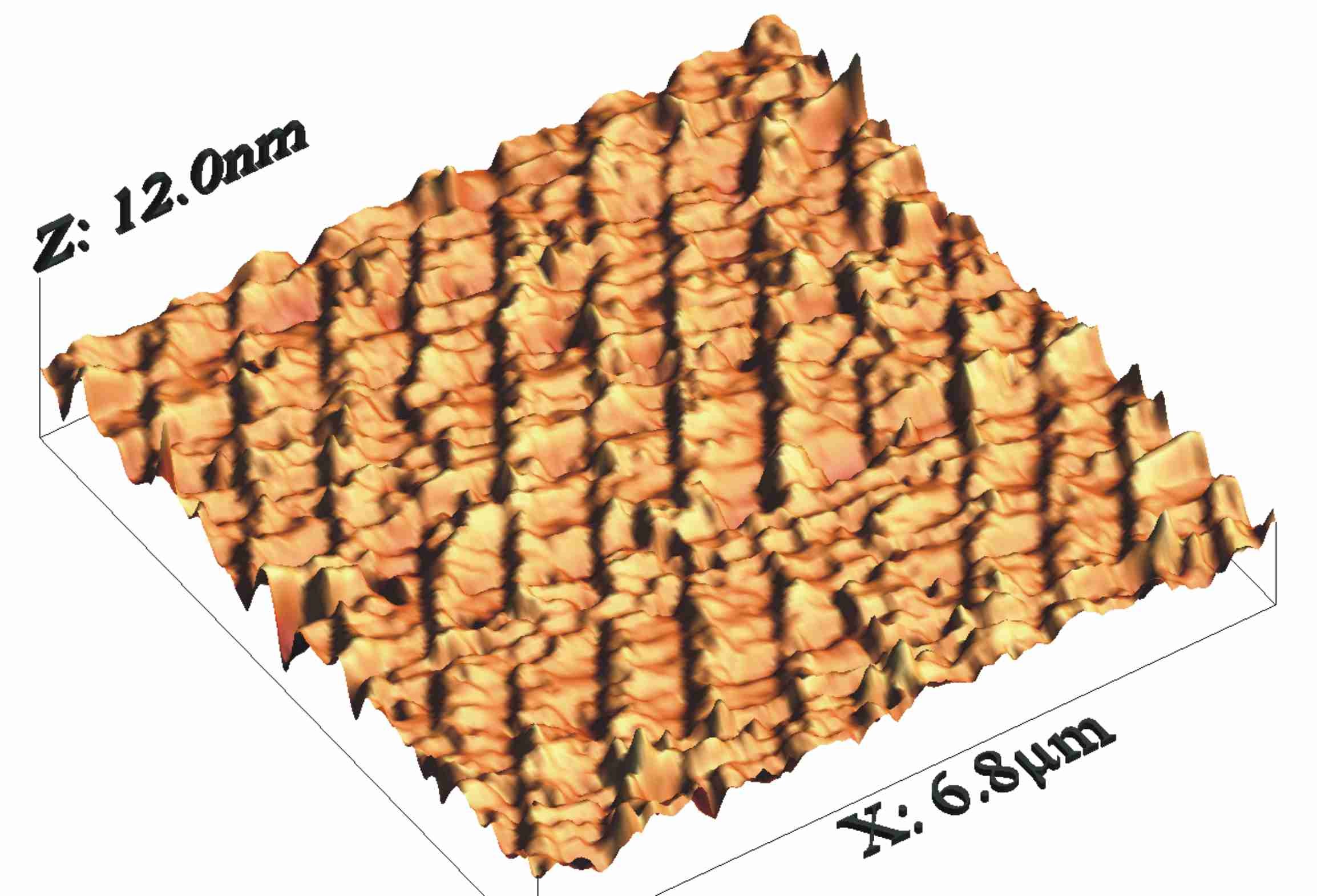}
\caption{Room temperature contact-mode AFM image of a NEG-123 single
crystal. The line-shaped features run along [100] and [010]
direction.\label{FigRTSTM}}
\end{figure}
The NEG-123 single crystals were grown from self-flux using oxides
and BaCO$_3$ with purity better than 4N. Growth took place in a
ZrO$_2$/Y crucible in an atmosphere of 100 mbar of air. After growth
the crystals have been oxidized in 1 bar of O$_2$ at 698 K for 140 h
and then at 667 K for 160 h. The crystals were characterized by
X-ray diffraction. The unit cells are orthorhombic and the crystal
is twinned. The transition temperature ($T_c$) as determined from
the $dc$-magnetization measurements is 93.14 K (mid point of
transition) with $\Delta T_c <$ 3 K.
\begin{figure}
\begin{center}
\includegraphics[width=0.4\textwidth]{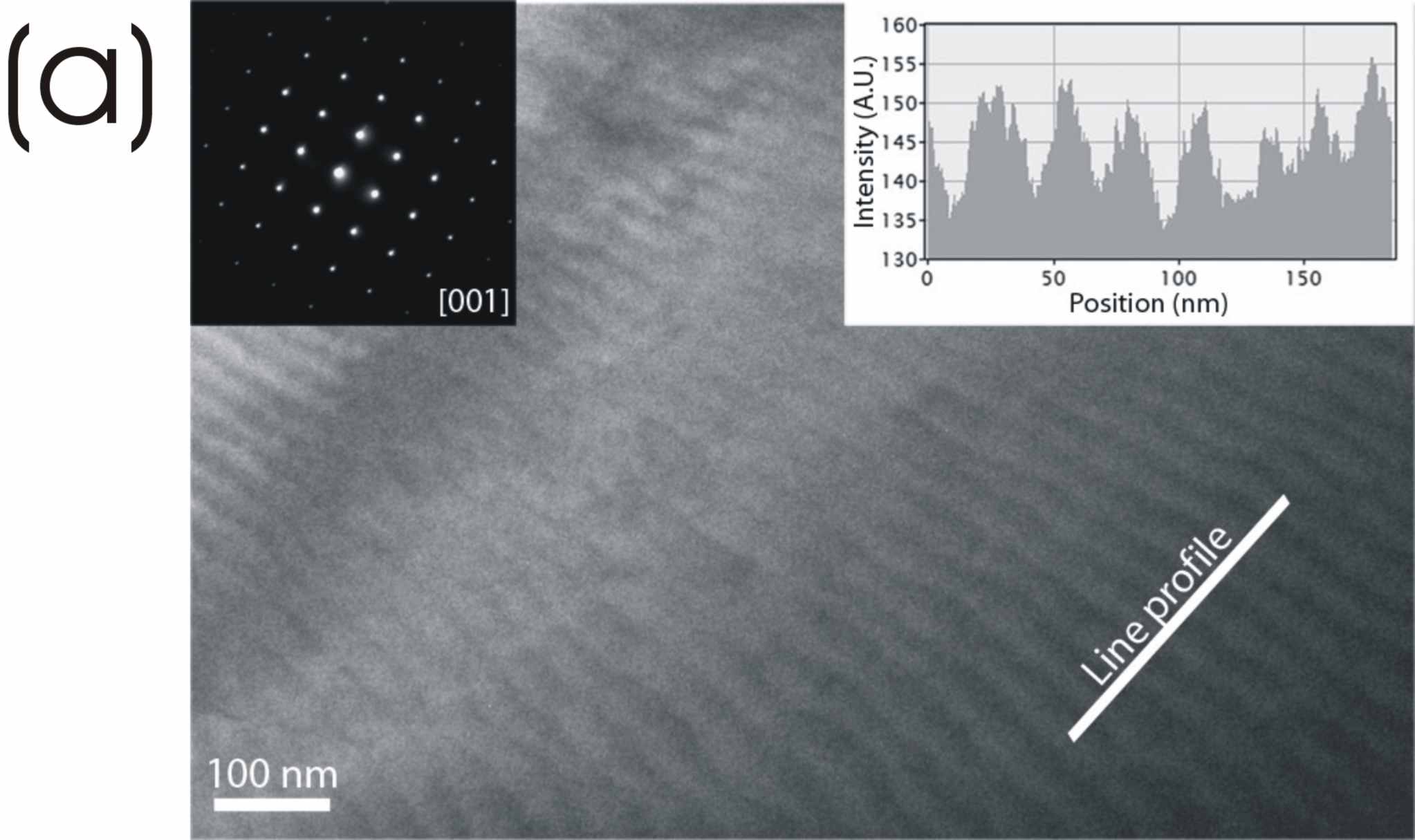}
\includegraphics[width=0.4\textwidth]{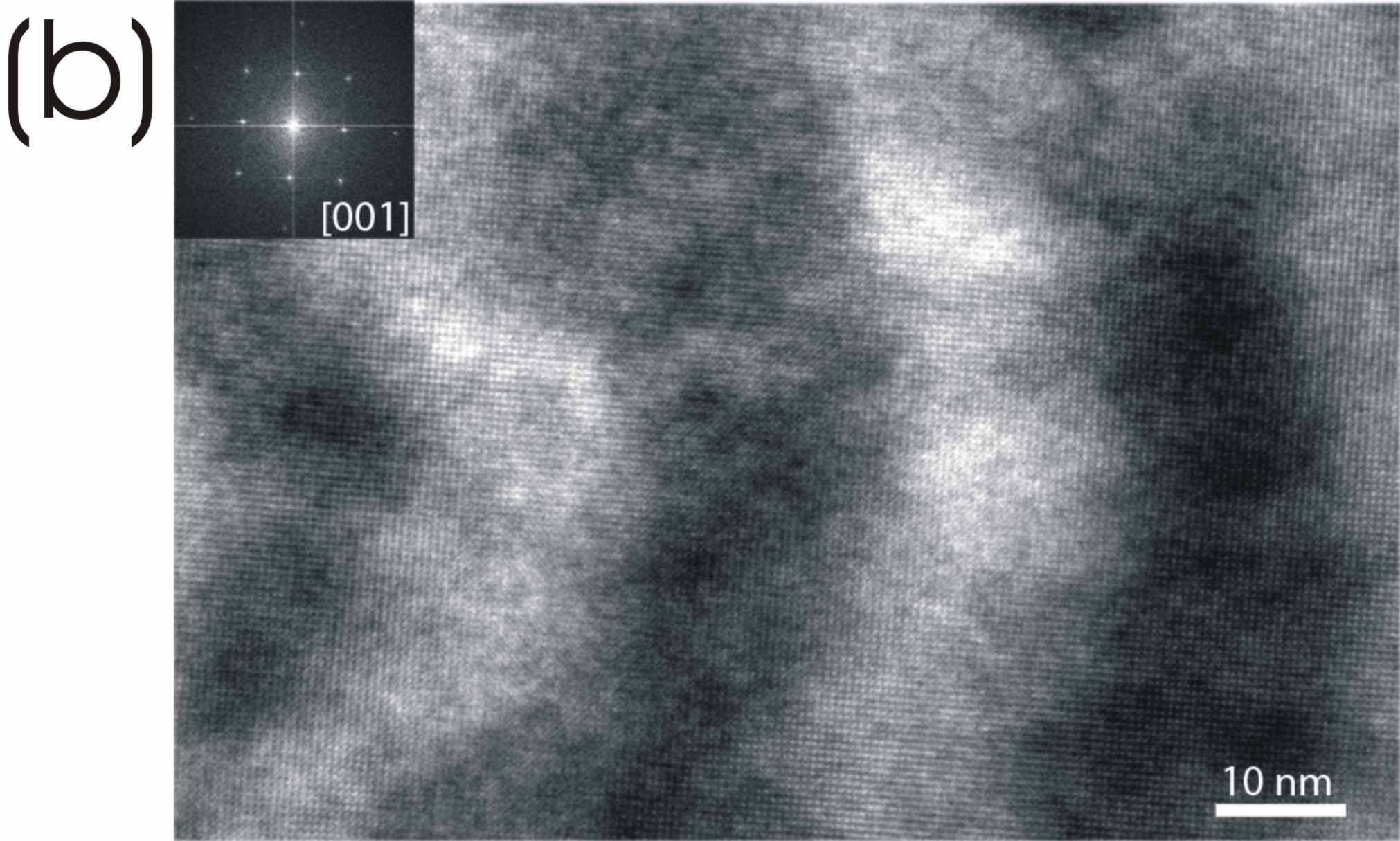}
\includegraphics[width=0.4\textwidth]{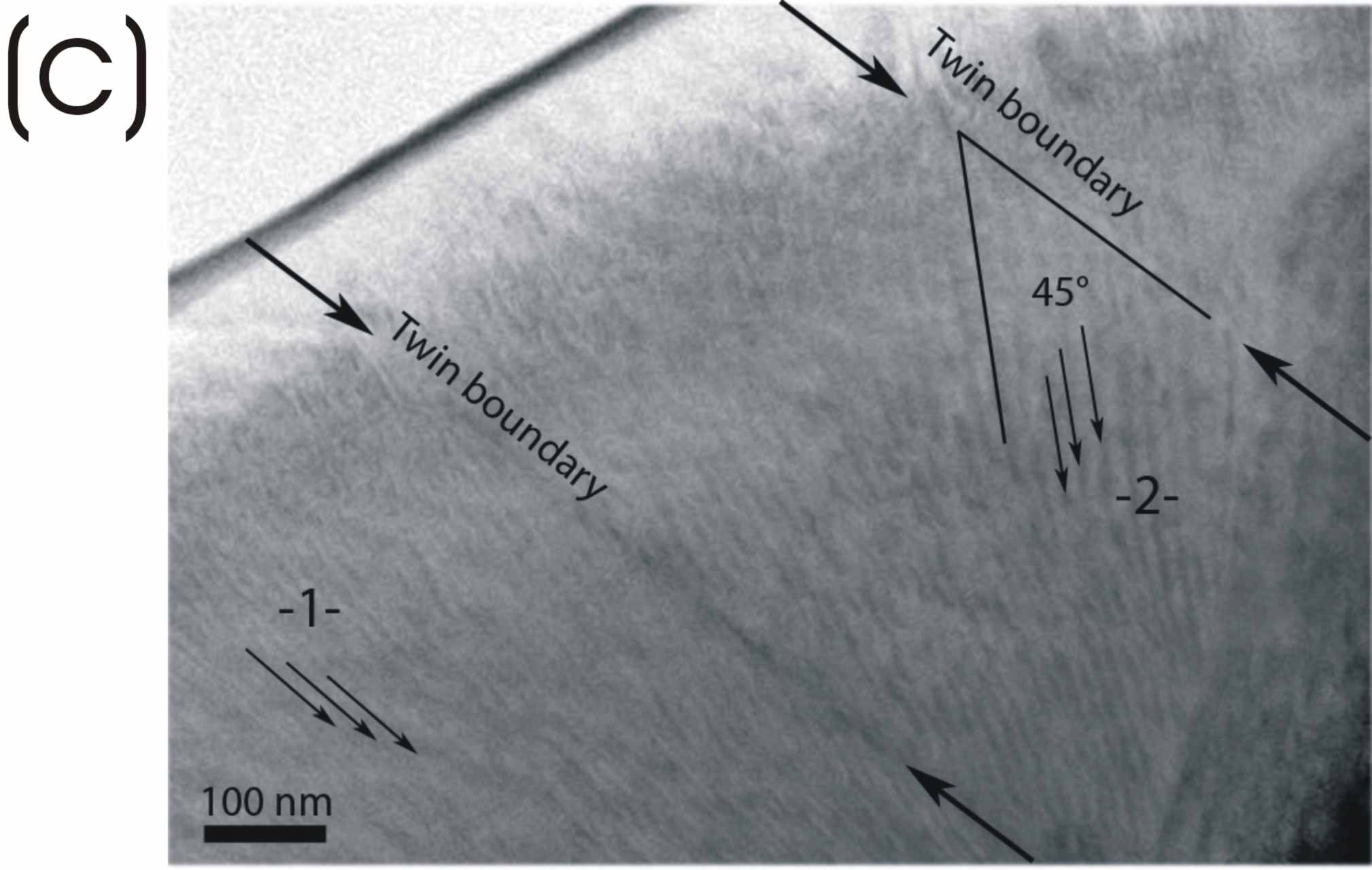}
\caption{(a) Typical bright-field TEM image taken from the NEG-123
single crystal sample along [001] zone axis. The line-profile shows
stripes of periodicity $\sim$30 nm (see inset). (b) High-resolution
image of a typical nanostripe region showing the lamellar-like
nature of the stripes. The [001] zone axis is evidenced by the inset
FFT. (c) TEM image showing the nanostripe orientation. In zone -1-
the nanostripes run almost parallel to the twin boundary. In zone
-2- the stripes run at a 45 degree angle with respect to the twin
boundary. \label{FigTEM}}
\end{center}
\end{figure}

Magnetization measurements were carried out at the field sweep rate
of 120 mT/s using a commercial (Quantum Design, PPMS) vibrating
sample magnetometer. $J_c$ was calculated from the M-H loops using
the extended Bean's model formula for a sample with rectangular
cross-section~\cite{Chen1989}. In Fig. \ref{FigJc}, $J_c$ at 65 K,
77 K and 84 K is plotted as a function of the applied magnetic
field. The curves exhibit a typical fish-tail shape with strong
secondary peak. With increasing temperature this peak shifts towards
lower field~\cite{Jirsa2007}. The secondary peak gives an evidence
of an effective random point-like disorder (active at intermediate
fields), similar to that in melt-processed LRE-123 samples. The
solid solution of RE$_{1+x}$Ba$_{2-x}$Cu$_3$O$_y$ and oxygen
vacancies are the sources for this type of
pinning~\cite{McCallum1995}. Apart from the secondary peak, another
significant feature of the curves is the central peak at zero field,
reaching $\approx$ 52 kA/cm$^2$ at $T$ = 77 K. For the secondary
peak even more than 62 kA/cm$^2$ is observed at this temperature.

In order to identify possible pinning sites, we performed AFM
studies on the as-grown sample surface. The measurements were
performed at room temperature and under ambient conditions with a
commercial AFM/STM (Nanoscope IV, Digital Instruments). As shown in
Fig. \ref{FigRTSTM}, a stripe-like structure is observed. The
periodicity of the stripes is in the range of 500 - 800 nm, which is
an order of magnitude larger than observed for melt-processed
samples \cite{Muralidhar2002, Koblischka2007, Koblischka2006}. The
profile is not of lamellar form as was observed on melt-processed
samples~\cite{Koblischka2006}.

\begin{figure}
\begin{center}
\includegraphics[width=0.4\textwidth]{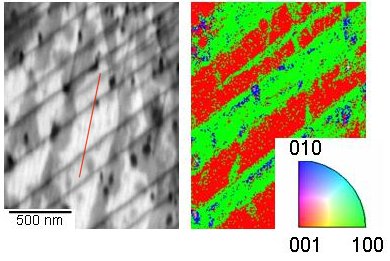}
\caption{(Color online) IQ map of the NEG-123 single crystal
obtained using EBSD (left). The line shows the direction of the
stripes. The corresponding IPF map (right figure) along the [001]
direction. The color code of the IPF map is given in the
stereographic triangle. \label{FigEBSD}}
\end{center}
\end{figure}

In order to elucidate if the stripes are related to vortex pinning,
the microstructure in the bulk was studied by TEM using a JEOL 3000
F microscope equipped with EELS and EDX facilities. The sample was
prepared by Ar ion milling. The measurements were performed along
the [001] zone axis. Figure \ref{FigTEM} shows a bright field image,
exhibiting a typical pattern observed by TEM. Stripe-like features
have been observed in the entire sample. The average spacing ranges
from 20 nm to 40 nm (see inset of Fig. \ref{FigTEM}(a)). A typical
high resolution image of the stripes is shown in Fig.
\ref{FigTEM}(b). The inset fast fourier transform (FFT) shows the
[001] zone axis and crystal orientation. The stripes are found to be
parallel to the twin boundaries in one area (-1- in Fig.
\ref{FigTEM}(c)) and inclined at an angle of 45 degrees with respect
to the twin boundaries in another (area -2- in Fig.
\ref{FigTEM}(c)). Thus, no clearly preferred orientation of the
stripes with respect to the twins is observed. From the periodicity
and topographic profile, the stripes observed in the bulk are
clearly different from those observed on the surface. The latter
were found to be parallel to one of the crystal edges thus being
most likely growth steps. The growth-steps, generally present only
on the surface~\cite{Lozanne1999}, have very weak pinning
performance compared to the defects in the bulk (e.g.,
twins)~\cite{Maple2000}. Important additional crystallographic
information is obtained from the image quality (IQ) map of the
recorded Kikuchi pattern at each point of the sample, obtained using
EBSD~\cite{Koblischka2007b}. The EBSD setup (TSL) is based on a FEI
dual-beam workstation (see Ref. \cite{Koblischka2007} for details).
Figure \ref{FigEBSD} shows the IQ map of a part of the sample and
its corresponding inverse pole figure (IPF) map which reveals the
crystallographic orientation. The IPF map shows that there is no
crystallographic misorientation due to the nanostripes as observed
for melt-processed samples~\cite{Koblischka2007}.

Hu \textit{et al.} suggested the stripes to originate from eutectic
growth~\cite{Hu2003}. Kirkaldy showed that during growth, a lamellar
structure could be produced from an eutectic solution having two
different phases~\cite{Kirkaldy1984}. However, in the present case,
crystal growth was stopped far above the eutectic temperature. Thus
the proposed mechanism could not apply here. It is not known if the
stripes are formed during growth or during oxygenation. The origin
is very likely to be material-dependent, where the composition
fluctuation plays a role, which is possibly specific to these kinds
of binary or ternary cuprate superconductors.

\begin{figure}
\begin{center}
\includegraphics[width=0.5\textwidth,height=50mm]{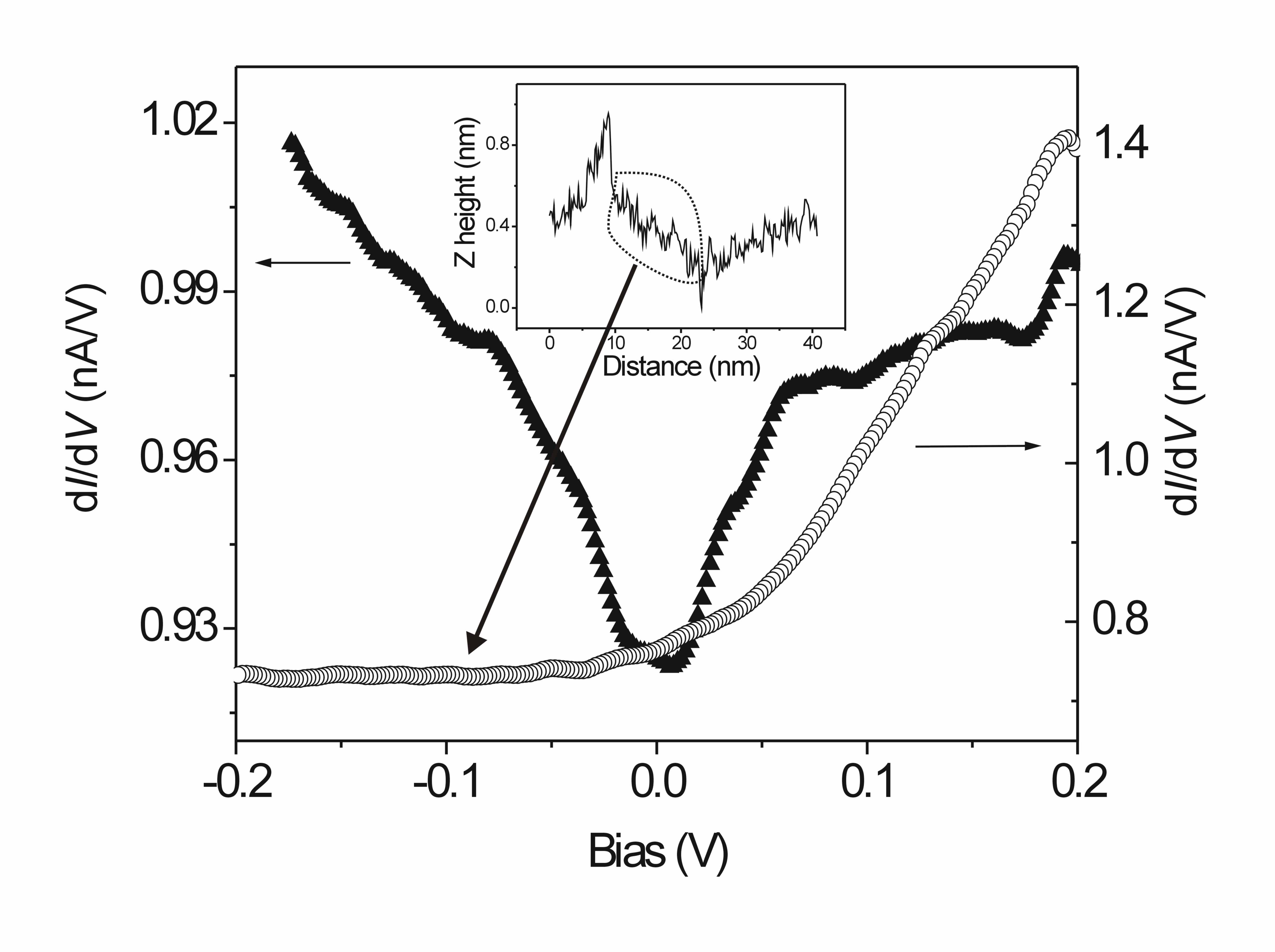}
\caption{Spectra across a twin boundary. The inset shows a cross
section of the boundary. Unfilled circles mark a typical curve
observed on the shoulder (indicated by an arrow). The curve shown by
filled triangles is representative for the entire superconducting
matrix.\label{FigLT}}
\end{center}
\end{figure}

Scanning tunneling spectroscopic (STS) experiments were performed on
the sample surface using a home-made STM~\cite{Zhang2001} under UHV
conditions. Apart from the growth steps, the surface of the crystals
also contains twins. These defects run through the crystal along
[110] and [1$\bar{1}$0] directions. The local electronic properties
across the twins were studied in the superconducting state ($T$ = 10
K). Representative STM and STS data are shown in Fig. \ref{FigLT}.
Spectroscopic data were taken at many different locations along and
close to the twin boundaries. d$I$/d$V$ curves obtained on the
exposed part of the boundary are similar to the curves obtained on
the entire area. These curves indicate a gap opening across the
Fermi level. The curves are slightly asymmetric with the filled
state gap edge higher than that of empty ones. The asymmetry of
spectra representing superconducting areas has often been
observed~\cite{Anderson2006, Das2007}. In contrast, the d$I$/d$V$
curves from the shoulder area are almost flat across the Fermi
level. They are highly asymmetric with empty sample state
conductivity higher than filled state ones. The underlying
composition in this region is not known. It is, however, evident
that the sample is locally nonsuperconducting. The width of the
normal conducting region is of the order of 2$\xi$. The absence of
superconductivity in these regions could be due to a different
oxygen content close to the twin boundary as compared to the entire
superconducting matrix. The twin boundaries represent regions of
high stress~\cite{Maple2000}. We note here that this is the first
direct observation of nanometer sized nonsuperconducting region
along the twin boundaries of a sample in the superconducting state.
So far the pinning effect of such nonsuperconducting regions is not
clear. From the magnetization measurements, we have not observed any
signature of pinning by these non superconducting regions. These
regions might guide the vortex motion along the direction of the
twin boundaries leading to channeling of vortices. The channeling
effect was observed as a depression of the secondary peak in $J_c-B$
curves obtained for this sample \cite{Jirsa2007}.

In summary, we observed stripes of periodicity of a few hundreds of
nm on the surface of NEG-123 single crystals. These stripes are most
likely the growth steps and are present only on the sample surface.
Stripes with periodicity of a few tens of nm were observed in the
bulk. These bulk defects do not exhibit any preferred orientation in
the crystal. Although no clear composition fluctuation could be
detected across the stripes, we believe that the stripes are
important for strong pinning similar to those in the melt-processed
samples. Low temperature STM/STS data showed nonsuperconducting
areas of nanometer-scale width aligned with the twin boundaries. The
potential of these nonsuperconducting regions towards vortex pinning
so far remains unclear.

\end{document}